\documentclass{optica-article}

\journal{opticajournal} 

\articletype{Research Article}

\usepackage{lineno}

\usepackage{graphicx}%

\begin{document}

\title{Polarization resolved hyperspectral imaging of the beetle \textit{Protaetia speciosa jousselini}}

\author{Peyman Soltani,\authormark{1,*} Milan ten Hacken,\authormark{1}, Arie van der Meijden,\authormark{2}, Frans Snik,\authormark{3} and Michiel J.A. de Dood\authormark{1,$\dagger$}}

\address{\authormark{1}Huygens-Kamerlingh Onnes Laboratory, Leiden University, Niels Bohrweg 2, Leiden, 2333 CA The Netherlands\\
\authormark{2}BIOPOLIS Program in Genomics, Biodiversity and Land Planning, CIBIO, Campus de Vairão, 4485-661, Vairão, Portugal\\
\authormark{3}Leiden Observatory, Leiden University, Einsteinweg 55, Leiden, 2333 CC The Netherlands}

\email{\authormark{*}soltani@physics.leidenuniv.nl, \authormark{$\dagger$}dood@physics.leidenuniv.nl} 


\begin{abstract*}
Understanding and classifying chiral structural color of scarab beetles across the phylogenetic tree is an important scientific tool to explore the origins of the homochiral optical response in biological structures with a potential relation to the homochirality of (chitin) molecules. We report hyperspectral polarization resolved images of the scarab beetle \textit{Protaetia speciosa jousselini} (Gory \& Percheron, 1833) and resolve the state of polarization as a function of both position (spatial resolution of $sim$20 $\mu$m $\times$ 10 $\mu$m) and wavelength (spectral resolution of 5.5 nm). We observe a strong left-handed chiral Bragg reflectance and analyze the center wavelength and width of the spectrum. The reflectance and state of polarization scale with the center wavelength of the Bragg reflection, while the relative width of reflectance spectrum is characterized by a chiral photonic strength parameter $\psi_C \approx 0.14$  that is independent of the reflected color and position. Based on the self-similarity of the reflectance spectra we interpret variation in reflectance as variations in the thickness of the chiral Bragg reflector across the beetle. 
\end{abstract*}

\section{Introduction}\label{sec1}

Reflected, transmitted and/or scattered light from biological structures contains valuable information about the underlying structure. This information is encoded in the wavelength, angle and polarization dependence of the reflectance. Pigments create color by absorption while wavelength-scale structures cause interference, diffraction and scattering effects leading to structural color~\cite{seago2009gold}. Unlike absorption-based pigments, structural colors are iridescent and depend strongly on angle of incidence. Both interference and interface reflections are polarization- and angle-dependent and thus add an additional layer of information to optical measurements. 

To resolve the wavelength, angle- and polarization-dependent reflectance, we built a polarization-resolved hyperspectral imager to investigate structural color. In comparison to standard color cameras that resolve red, green and blue images our hyperspectral imager resolves 240 simultaneous images at different wavelengths across the visible spectrum. In particular, we investigate the scarab beetle \textit{Protaetia speciosa jousselini} (Gory \& Percheron, 1833). The cuticle of this scarab beetle shows strong polarization- and wavelength-dependent reflectance resulting in brilliant green, red and blue colors. Scarab beetles are known to reflect left circular polarized light related to the prefered "handedness" of the chiral chitin structure~\cite{goldstein2006polarization}. Scarab beetles can see polarized light\cite{Warrant2010}, but there is no evidence for a behavioral response to polarized light\cite{Blaho2012}. A study of the wavelength-dependent chiral response of beetles may help to classify beetles according to their optical response and may help to understand the role of genetic information by mapping the optical response onto a phylogenetic tree~\cite{carter2016}. 

Natural chiral structures, such as those found in scarab beetles, have fueled a field of chiral photonics where artificially created chiral structure generates color that depends on the circular polarization state of light. Macroscopic homochiral structures appear in many biological systems but the origin of this homochirality is not completely understood~\cite{Devinsky2021, Robert2023How}. Homochirality of the optical response of the macroscopic structure could be linked to the homochirality of life's molecular building blocks~\cite{neville1969scarabaeid} through self-assembly of chiral molecules found in nature, e.g. left-handed amino acids, right-handed (poly-)saccharides, and the chiral double helix structure of DNA. Prominent, macroscopic chiral (nano)structures are found in some plants, fruits, and the exoskeletons of specific beetles~\cite{lininger2023chirality} that use either cellulose or chitin as a molecular building block. 
 
In this work we focus on the cuticle of jewel scarabs that consists of twisted layers of chitin nanofibrils where the periodicity of the structure matches the wavelength of visible light. These natural structures selectively reflect left-handed circularly polarized light within a specific wavelength band, centered around a central wavelength,~\cite{lowrey2007observation}. The origin of the colors is interference-based~\cite{rayleigh1919vii} and several studies report circularly polarized luminescence from chiral structures~\cite{sang2019circularly}.

The cuticle of the \textit{Chrysina} genus beetles has been extensively studied for its optical properties ~\cite{finlayson2017optically,mendoza2018graded, michelson1911lxi, neville1969scarabaeid,caveney1971cuticle,goldstein2006polarization,hodgkinson2010mueller, Arwin:13, Arwin:15, Arwin:16}. Seago et al. classified the structural color of beetles into multilayer reflectors, three-dimensional photonic crystals, and diffraction gratings~\cite{seago2009gold}. Different studies have reported wide variations in the handedness and spectral response of circularly polarized (CP) light reflections in \textit{Chrysina resplendens}. Ellipsometry confirms that scarab beetles predominantly reflect left-circular polarization (LCP) at visible wavelengths, with some beetles exhibiting elliptical polarization and right-handedness at red and near-infrared wavelengths~\cite{goldstein2006polarization}. Pye examined the phenomenon of CP light reflectance in many groups of scarabs without distinguishing wavelength-dependent effects~\cite{pye2010distribution}. Hodgkinson et al. used ellipsometric methods to explore \textit{Chrysina resplendens} and found both right-circular and left-circular polarizers in the 400 to 900 nm wavelength range~\cite{hodgkinson2010mueller,fernandez2011investigation}. Mendoza Gálvan et al. used electron microscopy images to model optical properites of \textit{Chrysina resplendens}, showing both right- and left-handed circularly polarized reflectance in a complex multilayer homochiral structure~\cite{mendoza2019mueller}.

Researchers have proposed chiral structural coloring mechanisms involving chiral photonic crystals and mixed media films. For example, chiral layers threading through each other in \textit{Pyronota festiva} form photonic crystals ~\cite{de2005natural}, and perturbations in these structures can shift polarization states ~\cite{hodgkinson2004threaded}. These findings, combined with nanoengineered proof-of-concept films and the structural complexity of insect cuticles, suggest that scarab beetles may inspire future developments in chiral photonics~\cite{hodgkinson2005structurally}.

The origin of this chiral polarization response is believed to be the twisted Bouligand structure in the exoskeleton, which can be visualized through electron microscopy and distinguished by specific sample preparation techniques ~\cite{lenau2008colours}. Other potential structures include molecular chirality and chiral photonic crystals observed in both beetles and butterflies ~\cite{schroder2011chiral,saba2011circular}. Recent work by Bagge et al. on the circularly polarized reflectance from golden scarab beetles has continued the exploration of these phenomena~\cite{bagge2020mueller}.
Advanced characterization techniques, including scanning electron microscopy (SEM) and  atomic force microscopy (AFM), have been employed to study these structures in detail. Additionally, polarization spectroscopy methods akin to Mueller-matrix measurements have been used to investigate the polarization properties of these beetles, although the full potential of this technique has not yet been exploited. 

The \textit{Protaetia speciosa jousselini} beetle serves as an exemplary case of natural chiral structures, displaying distinct structural coloration. Literature that reports images of \textit{Protaetia speciosa jousselini} and similar beetles observed through left- and right-circular polarizers are sparse~\cite{hegedus2006imaging} and are limited to the red, green and blue pixels of color CCD cameras. The observations through a right-circular polarizer suggest that these beetles exhibit reduced signal intensity. To further investigate the spatial distribution of polarization states we introduce a Hyperspectral-Stokes imager (HSSI) that records a polarization and wavelength-resolved datacube of images of the beetle's cuticle. The data, collected at various angles of incidence allows to assess the structural colors of the blue, green, and red segments of the cuticle of \textit{Protaetia speciosa jousselini}.

\section{Materials and Methods}\label{sec2}
Stokes polarimetry is a well-established optical technique for analysis of the state of polarization (SoP) of light, and plays a crucial role in, among others,  spectroscopy. The SoP of light is quantified by four Stokes parameters: \(S_0\) (total intensity), \(S_1\), \(S_2\), and \(S_3\) (components of the polarization state). A straightforward and intuitive method for measuring these parameters involves determining the intensities of four polarized light components: \(I_H\) (horizontal), \(I_D\) (diagonal), \(I_A\) (anti-diagonal),\(I_R\) (right-circular), and \(I_L\) (left-circular).

\begin{figure}[htb]
\centering
\includegraphics[width=0.6\columnwidth]{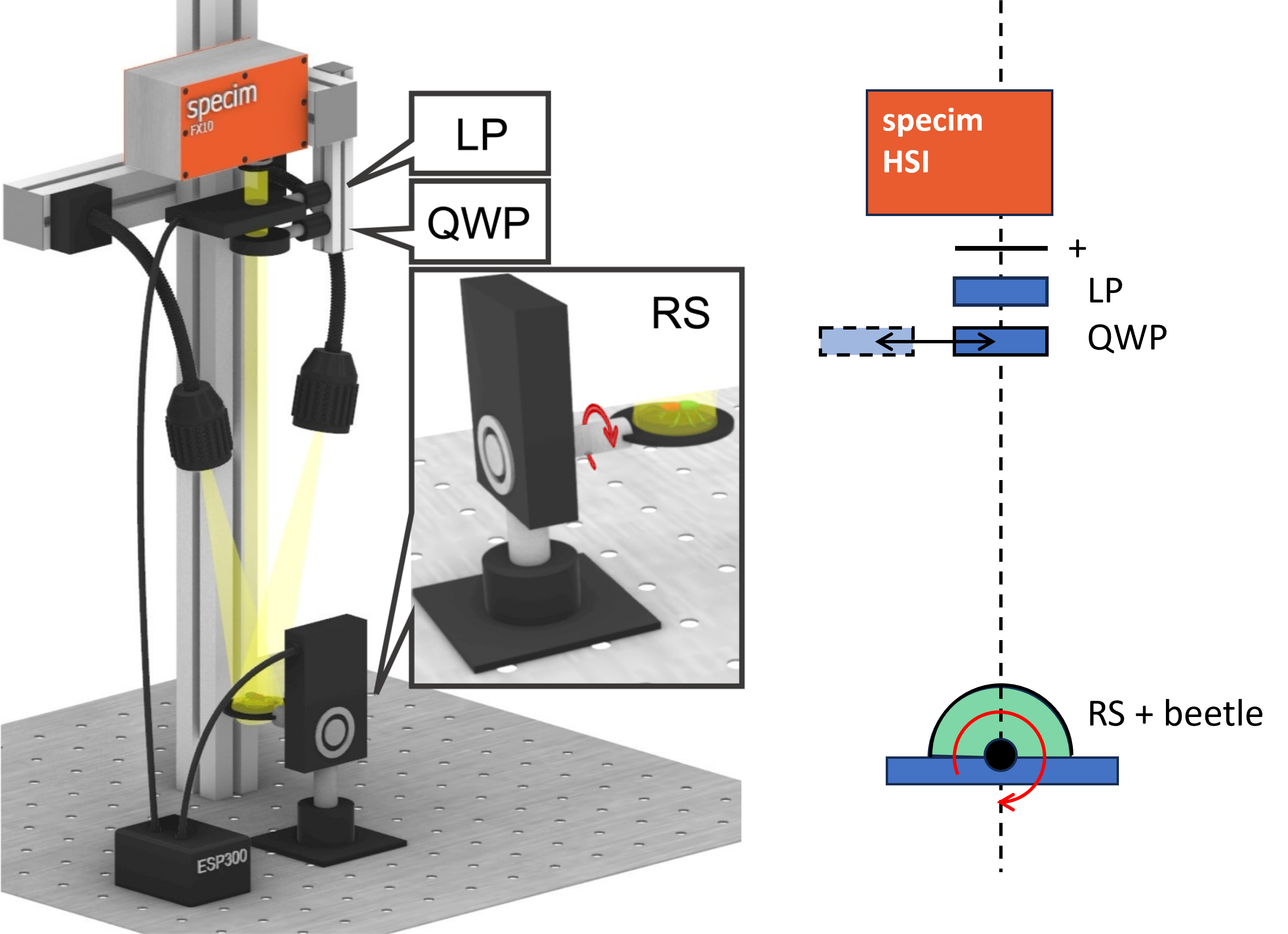}
\caption{Schematic of the hyperspectral Stokes imager. The quarter-wave plate (QWP), linear polarizer (LP), and roll-axis rotation stage (RS) are used to scan the beetle. The ais of the rotation stage is aligned with the slit of the camera.  Both the linear polarizer and the roll-axis rotation stage are rotated using a motorized rotation stage (Newport PR50PP). Two stable halogen lamps (RMS intensity fluctuations below $0.35\%$)  next to the camera are used to illuminate the beetle along the slit (the uniformity of illumination is $>98\%$). The center of the light beam corresponds to a 12° angle of incidence.
}\label{fig1} 
\end{figure}

These intensities are straightforwardly measured using a linear polarizer and a quarter-wave plate (QWP). The polarizer selectively transmits light polarized along its transmission axis, facilitating the measurement of linear polarization intensities, i.e. \(I_H\), \(I_D\) and \(I_A\), where we define the horizontal direction (0°) along the direction of the imaging slit of the hyperspectral camera. Inserting the QWP introduces a \(\pi/2\) phase shift, between orthogonal polarization directions. The intensity profiles \( I_R \) and \( I_L \) are obtained with the slow-axis of the QWP aligned with the  camera slit and the linear polarizer along the diagonal \(I_D\) ( 45°) and anti-diagonal \(I_A\) (135°) directions. The measurements are corrected for the transmittance of the QWP at different wavelengths, by introducing the factor $\tau^*$ as ratio of the intensities 
\begin{equation}
\tau^*=\frac{I_D+I_A}{I_R+I_L}
\end{equation}
The linear polarization intensities are measured  with the polarizer only set to 0° (\(I_H\)) , 45° (\(I_D\)) and 135° (\(I_A\)), respectively.

The Stokes parameters are then determined using the following equations~\cite{Singh2020, Snik2013, Snik2014}:
\begin{align}
S_0 &=  I_D + I_A \\
S_1 &=  2I_H - S_0 \\
S_2 &=  I_D - I_A \\
S_3 &=  \tau^* (I_R - I_L)
\label{eq1}
\end{align}

The experimental setup for obtaining spectrally resolved polarization images is illustrated in Fig.~\ref{fig1}. An achromatic quarter-wave plate at 580 nm (QWP, Thorlabs AQWP10M-580) and a linear polarizer (LP, PRINZ) are positioned in front of the hyperspectral imager (FX10, Specim Co., Inc.) to resolve Stokes parameters at various wavelengths. The hyperspectral Stokes imager (HSSI) operates as a line-scan camera with 1024 pixels and produces 240 spectral images in the wavelength range of 400-1000 nm (5.5 nm spectral resolution).

Using a motorized stage, the polarizer is automatically rotated to 0°, 45°, and 135°. For circular polarization measurments the fast axis of the QWP is aligned with the 90° direction. This configuration allows for the acquisition of the required intensity measurements. The limitations of both the polarizer and QWP limit the wavelength range of the HSSI to 450-800 nm.

We verify our HSSI by creating a known set of elliptical states of polarization. These states are created by rotating a second, identical achromatic quarter-wave plate in front of a linear polarizer at 0$^\circ$. We acquire the Fraction of Linear Polarization (FOLP) and Fraction of Circular Polarization (FOCP) for all wavelengths. The FOLP and FOCP are defined as~\cite{Singh2020, Snik2013, Snik2014}

\begin{align}
FOLP &= \frac{\sqrt{S_1^2+S_2^2}}{S_0}  \\
FOCP &= \frac{S_3}{S_0}
\label{eq2}
\end{align}

The standard deviation of these measurements was calculated to determine the maximum deviations of the camera over the 450-800 nm wavelength range. 

\begin{figure}[tb]
\centering
\includegraphics[width=0.7\columnwidth]{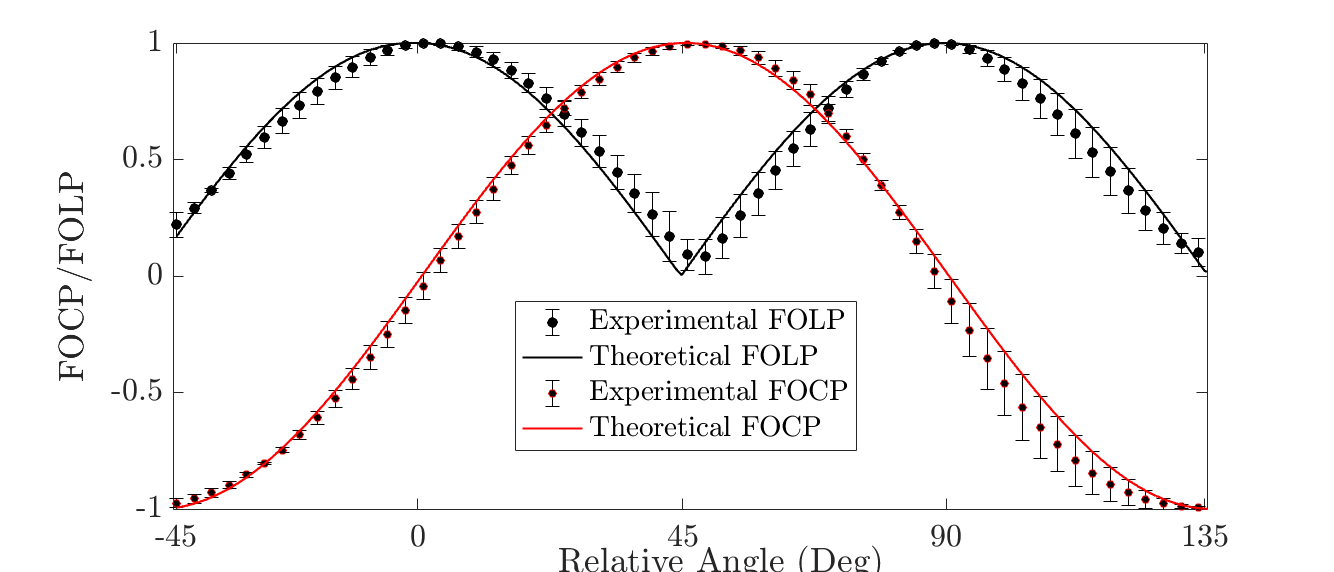}
\caption{Verification of the HSSI as a function of relative angle between the linear polarizer and QWP. FOLP (black) and FOCP (red) averaged over wavelength and spatial pixels of the HSSI. The solid lines represent theoretical values obtained by Jones calculus assuming a perfectly achromatic polarizer and quarter wave plate. These curves depend only on the relative angle $\Delta \varphi$ and are given by $FOLP(\Delta \varphi) = |\cos(\Delta \varphi)$ and $FOCP(\Delta \varphi)=\sin(\Delta \varphi)$ where $\Delta \varphi = 0$ corresponds to a linear polarization state aligned with the camera slit. The symbols with error bars illustrate a maximum 4\% deviation of the experimental values compared to the theoretical prediction (lines).
}\label{fig2}
\end{figure}

Fig.~\ref{fig2} illustrates the plot of the average Fraction of Linear Polarization (FOLP, black) and the Fraction of Circular Polarization (FOCP, red). The solid lines represent the theoretical FOLP and FOCP for a vertical polarization state rotated by an ideal achromatic quarter-wave plate.

The error bars indicate the deviations and show that the total deviation in FOLP and FOCP is less than 4\% for any wavelength when comparing to the calculation for ideal achromatic polarizers and QWPs. The reported deviation is the combined effect of generation and measurement of the polarization state. 

\section{Results and discussion}\label{sec3}

The HSSI is used to measure four specimens (3 male, 1 female) of the flower beetle \textit{Protaetia speciosa jousselini}. These beetles have shiny green elytra connected to the contrasting red triangular scutellum and neck shield (pronotum). The beetles are convex, oval, and slightly flattened with different radii of curvature when viewed from the side and front. The pronotum, located behind the head, is steeply convex, providing protection for the head and thorax. This morphology facilitates navigation through foliage and decaying wood but introduces challenges in measuring polarization and spectral properties of the angle dependent structural colors. The beetle's cuticle is a complex, multifunctional structure composed of multiple layers. The outermost layer, the epicuticle, is thin and waxy, serving as a barrier to water loss and microbial invasion. Beneath the epicuticle is the exocuticle, a rigid layer made of chitin fibers that form optically birefringent layers. These fibers in these layers twist to create periodic structures on a visible wavelength scale to generate the beetle's intricate coloration and polarization properties.
From a 3D scan of one beetle specimen we find a (mean) radius of curvature of 30.8 mm (side view), 11.0 mm (front view), and 9.5 mm (side view of the pronotum). 

To mitigate the influence of the beetle's frontal curvature on the measurements, we employ a roll-axis rotation stage with a radius of approximately 11 mm, matching the radius of the beetle's front view curvature. This setup allows to image the entire beetle by rotating the roll-axis rotation stage using the hyperspectral Stokes imager as a line-scan camera. A two-dimensional image is created from these line images where each line image is collected at near-normal incidence. 

\begin{figure}[h]
\centering
\includegraphics[width=0.9\columnwidth]{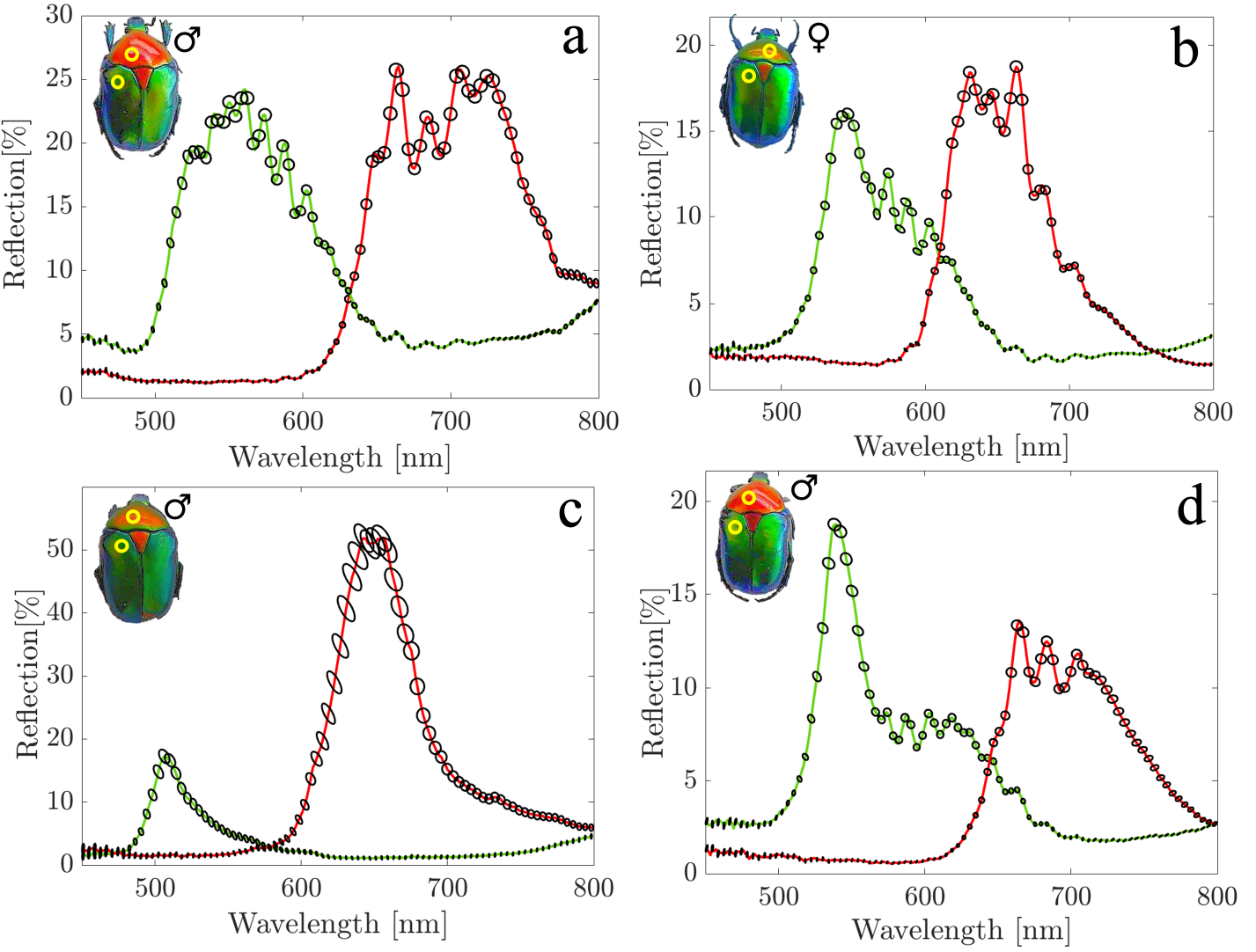}
\caption{Polarization resolved hyperspectral imaging of \textit{Protetia speciosa jousselini} beetles (a-d) Reflectance spectra of two distinct parts of the beetles (elytra, (green curves) and pronotum, (red curves)). The ellipses refer to the elliptical polarization state as a function of wavelength. The insets show images of each beetle with yellow circles indicating the locations where the spectra shown in the main figure are extracted.
}\label{fig3}
\end{figure}

Figure~\ref{fig3} shows the normalized reflectance spectra of the different specimens of the  \textit{ Protaetia speciosa jousselini} for the green and red colored parts (scutellum and elytra). As can be seen, these spectra display strong reflectance bands from the exocuticle structure. The ellipses in the figure show the polarization state of the reflected light demonstrating the strongly left-circularly polarized reflectance at the primary wavelength. The insets show color images of each beetle specimen.

\begin{figure}[htb]
\centering
\includegraphics[width=0.5\columnwidth]{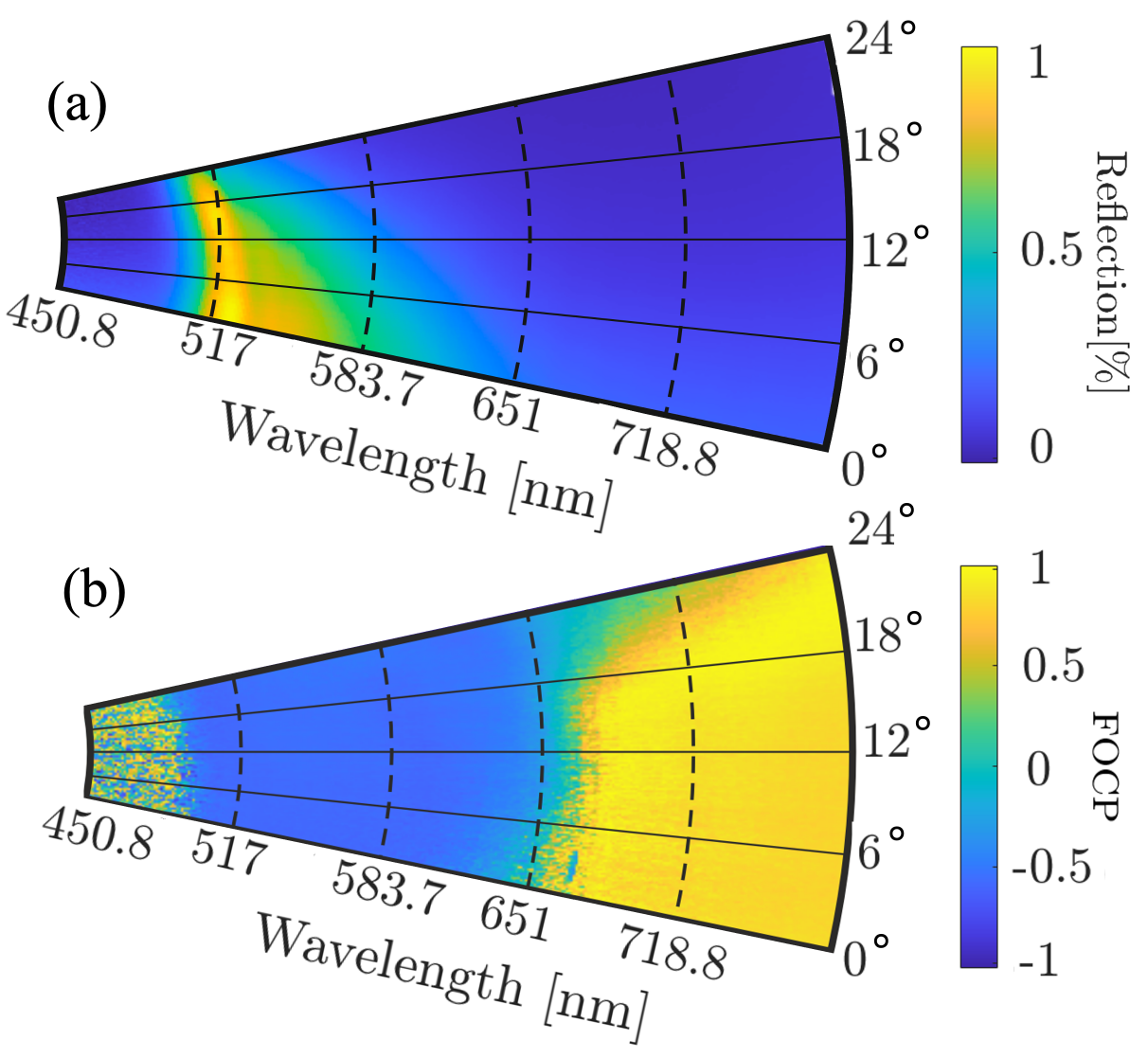}
\caption{Angle dependence of the optical properties of \textit{Protetia speciosa jousselini} False colour images of the reflectance (a) and Fraction of circular polarization (b) plotted as function of angle of  incidence and wavelength.}\label{fig4}
\end{figure}

Strong effects on the state of polarization are known to occur at angles close to 90°\cite{hegedus2006imaging}. To assess how the angle of incidence influences our measurements we measure the angle dependence of the spectral features close to normal incidence. A partially collimated halogen lamp with a 5 mm beam diameter was employed to uniformly illuminate a localized region of the beetle's green part. The lamp was mounted on a motorized rotation stage, allowing precise adjustment of the central incident angle from 0° to 24°, while the Hyperspectral Stokes Imager (HSSI) maintained a fixed observation angle of 20°. The accessible angular range is limited by the local curvature of the beetle. This configuration enabled measurement of the spectral and polarization characteristics of reflected light as a function of the incident angle and are consistent with data reported by Hegdus et al. \cite{hegedus2006imaging}. By measuring Stokes parameters and spectral data across varying angles, this experimental setup facilitated a comprehensive analysis of how the angle of incidence influences the polarization and spectral reflectance of the beetle surface.

The angle dependence, as illustrated in Fig.~\ref{fig4}, affects the measurement of the beetle using the HSSI. The light source in our HSSI experiments is centered at an angle of incidence of 12° with significant illumination at angles between 9° and 15°. This variation results in unwanted light due to scattering, specular reflections, and diffraction that reaches the HSSI and  contaminates the signal. From Fig.~\ref{fig4} we estimate that the angle of incidence in the HSSI induces a spectral blue shift up to 5.4 nm at the maximum deviation of 15°. We determine that the impact of the variation in angle of incidence on the fraction of polarization is less than 4\%. This value corresponds to the spectral range of 520 nm to 800 nm at 80\% of the camera's maximum detectable intensity for angles up to 15°. Additional errors due to misalignment and wavelength dependence of the achromatic QWP and polarizer also contribute less than 4\% to the total error (see Fig. \ref{fig2}). Statistical errors in the reflectance due to measurement noise and dark counts are comparable in magnitude. We thus estimate that our FOLP and FOCP data is typically significant at $\sim$10\% at any wavelength between 400 and 800 nm. An exception occurs in Fig.~\ref{fig4}b for wavelengths below 480 nm. The combined effect of low reflectance and relatively low signal results in larger fluctuations in the FOCP that are clearly visible in the figure. 

\begin{figure}[bth]
\centering
\includegraphics[width=0.9\columnwidth]{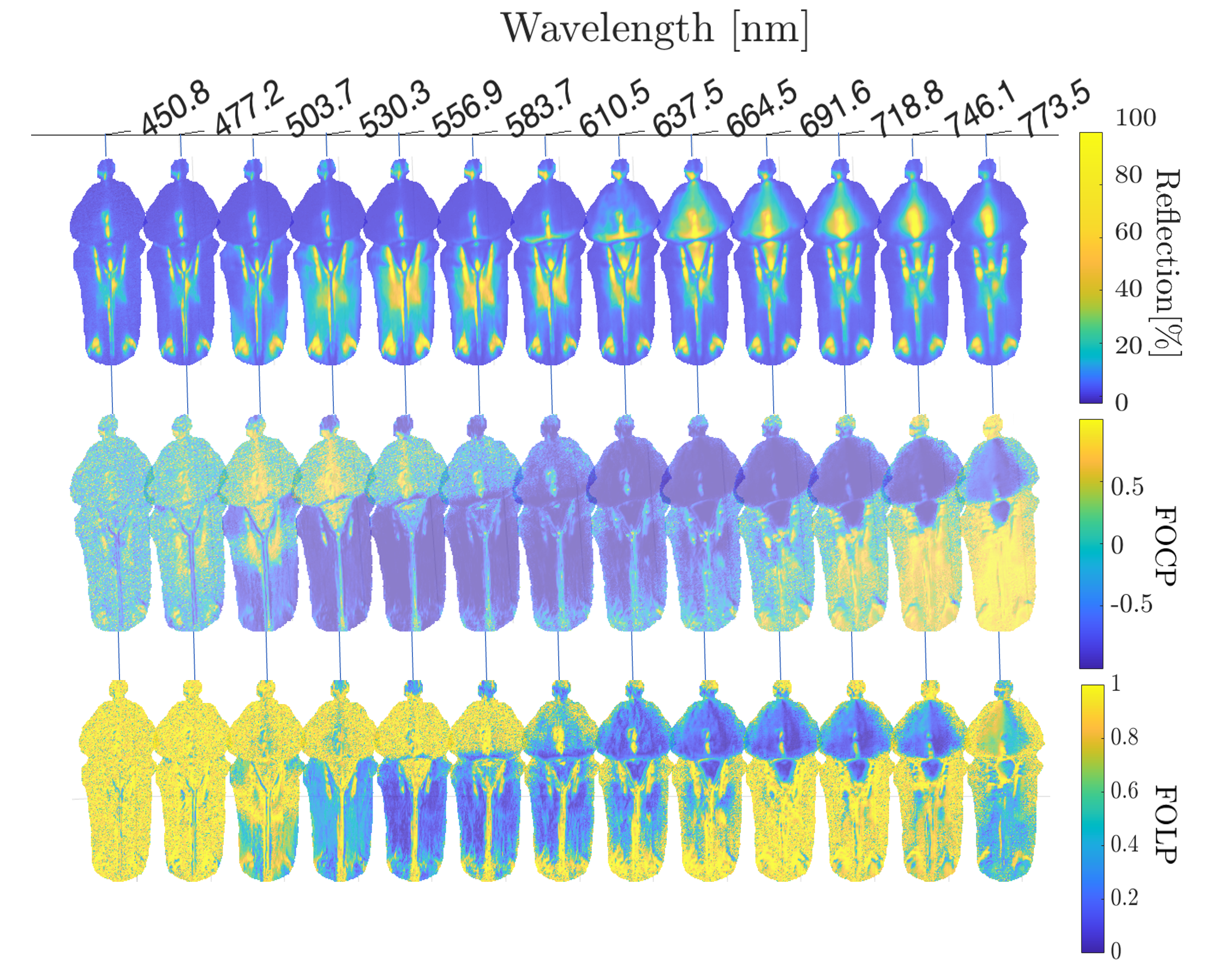}
\caption{Polarization resolved hyperspectral images of \textit{Protetia speciosa jousselini} showing (unpolarized) normalized reflectance (top row), Fraction of circular polarization (FOCP) (middle row), and Fraction of linear polarization (FOLP) (bottom row) images at selected wavelengths}\label{fig5}
\end{figure}

In the remainder of this article we will discuss the structural color mechanism of \textit{Protetia speciosa jousselini} at different locations by mapping different spectral and polarization properties over the surface of the beetle for the specimen presented in Fig.~\ref{fig3} (a). 
Fig.~\ref{fig5} presents a selection of hyperspectral images of the beetle. Images of the reflectance (top row), fraction of circular polarization (middle row) and fraction of linear polarization (bottom row) are shown for selected center wavelengths between 450.8 and 773.5 nm corresponding to the spectral pixels of the hyperspectral imager. As can be seen in the images in the top row, the reflectance of different parts of the beetle peaks at different wavelengths. 

The images in the middle row show strongly left-circular polarization (FOCP = -1) at the peak wavelength of the reflectance and showcases the chiral nature of the nanostructure for the green, red and blue parts of \textit{Protetia speciosa jousselini} as was originally reported by Hegedus et al.~\cite{hegedus2006imaging} using only the red, green and blue pixels of a standard RGB color camera. Linear polarization (bottom row) as opposed to depolarization (e.g due to scattering from surface roughness) is identified outside regions of strongly circularly polarized reflectance.

A closer inspection of our hyperspectrally resolved images reveals a  reversal in circular polarization indicated by the yellow color in Fig.~\ref{fig5} beyond $\sim$750 nm (middle row, rightmost images). This effect is reminiscent of the Cotton effect, where the FOCP reverses at wavelengths close to the edges of an absorption band. Around the maximum left-handed response (FOCP $\approx -1$), the circulair dichroism becomes zero and changes sign, an effect known as optical rotatory dispersion ~\cite{Karnik202195,zhang2022liquid}. 

The green regions (elitra) exhibit strong left-circular polarization in the wavelength range of 550 nm to 610 nm. Around 705 nm, the FOCP  is close to zero. For longer wavelengths, the elytra shows weak reflectance of right-circular polarization with the FOCP reaching approximately 75\%, at 770 nm. In addition, the red regions of the cuticle (pronotum and scutellum) show strong left-circular polarization around 580 nm and reversal of the FOCP at the blue side of the reflectance spectrum. 

We interpret this reversal as an effect due to a Fresnel reflection from the non-index matched surface of the beetle. While the bulk acts as a perfect chiral mirror that reflects left-circular polarization the surface does not. Hence, the reflectance of the beetle is in general elliptical. At the peak wavelength the bulk reflection is strong and close to left-circular. Outside the reflection band the Fresnel reflection is important and the combination with the bulk reflection gives a signal that is mostly right-circular polarized. 

The strong reflectance at a specific wavelength range combined with the circular polarization are due to a periodic chiral photonic structure in the beetles. The strong reflectance is due to constructive interference between the layers~\cite{lowrey2007observation}. Relevant parameters that influence chiral reflectance in these beetles include the refractive indices of the birefringent layers, the pitch of the chiral structure, the number of pitches, and the rotational angle between consecutive birefringent layers. This rotational angle can be estimated based on the number of chitin layers per pitch, e.g. by electron microscopy or atomic force microscopy~\cite{Yang2017}.

The different areas of \textit{Protetia speciosa jousselini} reflect light at different wavelengths, while the underlying mechanism and photonic structure might be similar. To analyze this self-similarity we introduce a chiral version of the photonic strength $\psi_C$~\cite{vos1996strong,shung1993surface,mccall2014birefringent}. This parameter measures the width of the spectrum of circularly polarized reflectance relative to the central frequency and is a measure of the index difference in the system. We define this photonic strength from the experimental data as 
\begin{align}
\psi_C &=  \frac{\Delta\omega_\text{Bragg}}{\omega_\text{Bragg}} \approx \frac{\Delta\lambda_\text{Bragg}}{\lambda_\text{Bragg}} \label{eq3}
\end{align}
where $\lambda_\text{Bragg}$ and $\Delta\lambda_\text{Bragg}$ are, respectively, the central wavelength and the spectral width of the peak in the reflectance spectra for circular polarized light. Here we assume that $\Delta\lambda_\text{Bragg} \ll \lambda_\text{Bragg}$. 

The theoretical basis of a chiral definition of photonic strength follows an analytical solution in a chiral medium with a continuous rotation of the anisotropic layers~\cite{Oseen1933, Kats1971, Belyakov1979, Lakhtakia1995} as a limiting case of the discrete rotation of layers in an ambichiral medium~\cite{hodgkinson2004ambichiral}. In this continuous approach a transformation~\cite{Oseen1933} to a rotating coordinate frame introduces the right and left circular polarized states as exact Eigenstates of the helical structure. Hence, the reflectance of an index-matched chiral structure is chiral and only circularly-polarized light that matches the handedness and pitch of the medium is reflected. Without perfect index matching the reflected light is elliptically polarized, but the reflectance spectrum of the index-matched chiral Bragg reflector and the corresponding photonic strength in the basis of left- and right-circularly polarized Eigenstates remains valid. 

A coupled wave analysis of a continuous chiral medium~\cite{mccall2014psi} results in a chiral Bragg wavelength that corresponds to the central wavelength in the circulary polarized reflectance spectrum. This wavelength is defined as 
\begin{equation}
\lambda_\text{Bragg} = 2 \bar{n} \Lambda
\end{equation}
where $\bar{n}$ is the effective average index and $\Lambda$ is the pitch of the medium. The width of the chiral reflectance is given by
\begin{equation}
\Delta \lambda = 2 \Delta n \Lambda
\end{equation}
where $\Delta n$ is the birefringence of the medium. The chiral photonic strength can thus be interpreted as a measure of the relative birefringence of the medium, i.e. normalized to the average refractive index.
\begin{equation}
\psi_C = \frac{\Delta n}{\bar{n}}
\end{equation}
We stress that a chiral photonic strength parameter $\psi_C$ can be obtained directly from the experimental data and represents a measure of the underlying structure of the different parts of the beetle. The fact that Maxwell's equations do not have an intrinsic length scale causes the entire reflectance spectra of self-similar structures to scale as $\Lambda/\lambda$, where $\Lambda$ is the pitch of the chiral structure. For self-similar structures that only differ in pitch the chiral photonic strength $\psi_C$ is expected to be independent of the central wavelength of the reflectance spectrum. 

\begin{figure}[h]
\centering
\includegraphics[width=0.8\columnwidth]{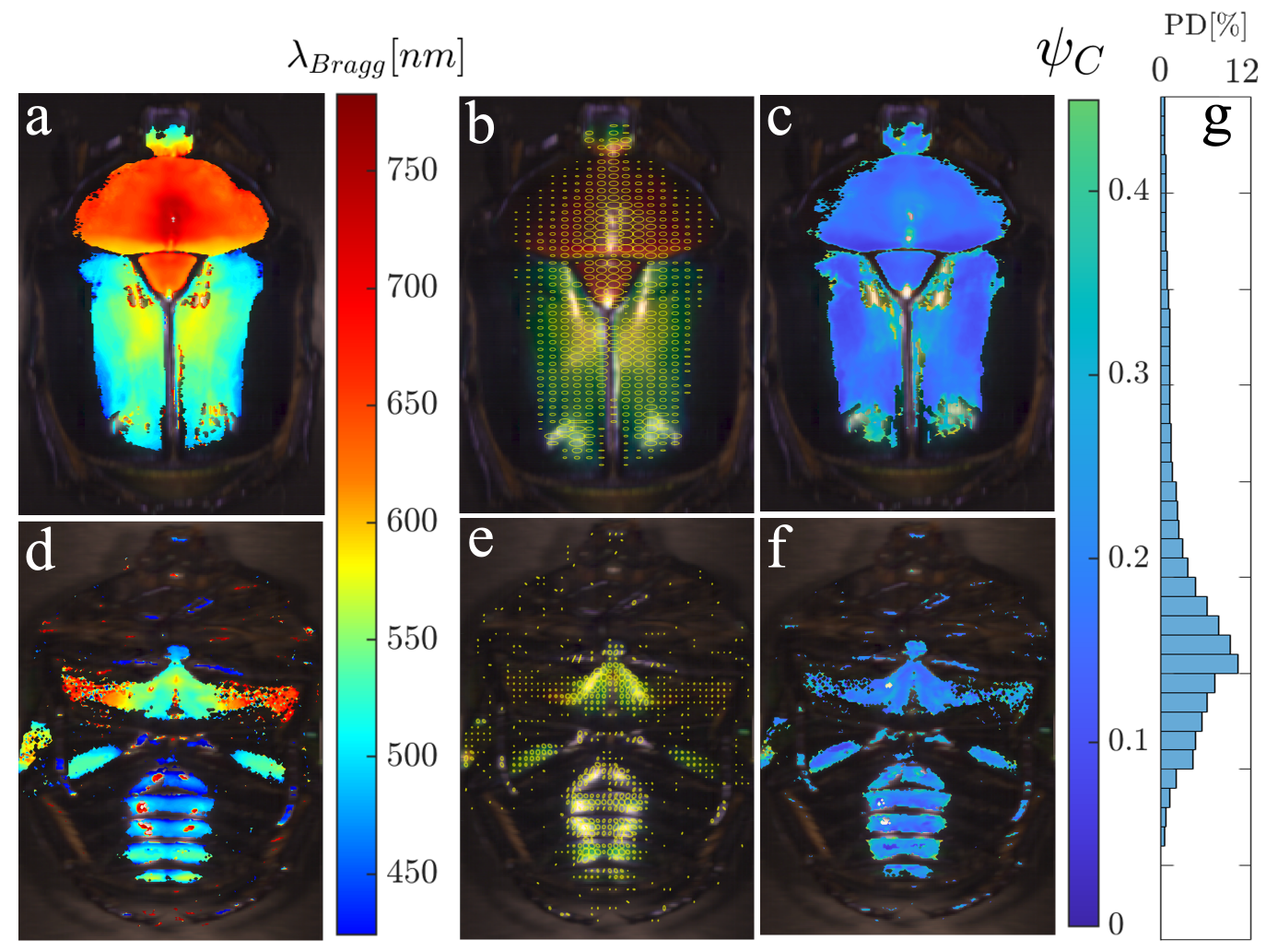}
\caption{Analysis of Bragg reflectance of \textit{Protetia speciosa jousselini}. Images are shown for the back (top row) and belly (bottom row). (a,d) False colour image of the Bragg wavelength (b,e) elliptical state of polarization at the Bragg wavelength superimposed on the beetle image, (c,f) false colour image of photonic strength, and (g) histogram of Pixel Density (PD), showing the percentage of pixels as a function of photonic  strength.}\label{fig6}
\end{figure}

We measured the polarized reflectance of the top and bottom of the abdomen of \textit{Protetia speciosa jousselini} as shown in Fig.~\ref{fig6}, revealing that they function as narrow-band, high-signal left-handed circular polarizers at specific wavelengths. The images (a) and (d) show the variation in the peak wavelength across the beetle, while (b) and (e) show the strong left-circular state of polarization at the central wavelength. The photonic strength is plotted in (c) and (f) and shows that the photonic strength parameter $\psi_C$ of \textit{Protetia speciosa jousselini} is mostly independent of the position on the beetle. This confirms a similar structural origin of the color and polarization properties for different regions on the beetle and indicates a self-assembly mechanism for scaled versions of identical chiral Bragg mirrors. Recent studies have also shown no significant differences in the chemical composition of the cuticle between these colors~\cite{lee2020structural}. 

\begin{figure}[h]
\centering
\includegraphics[width=0.7\columnwidth]{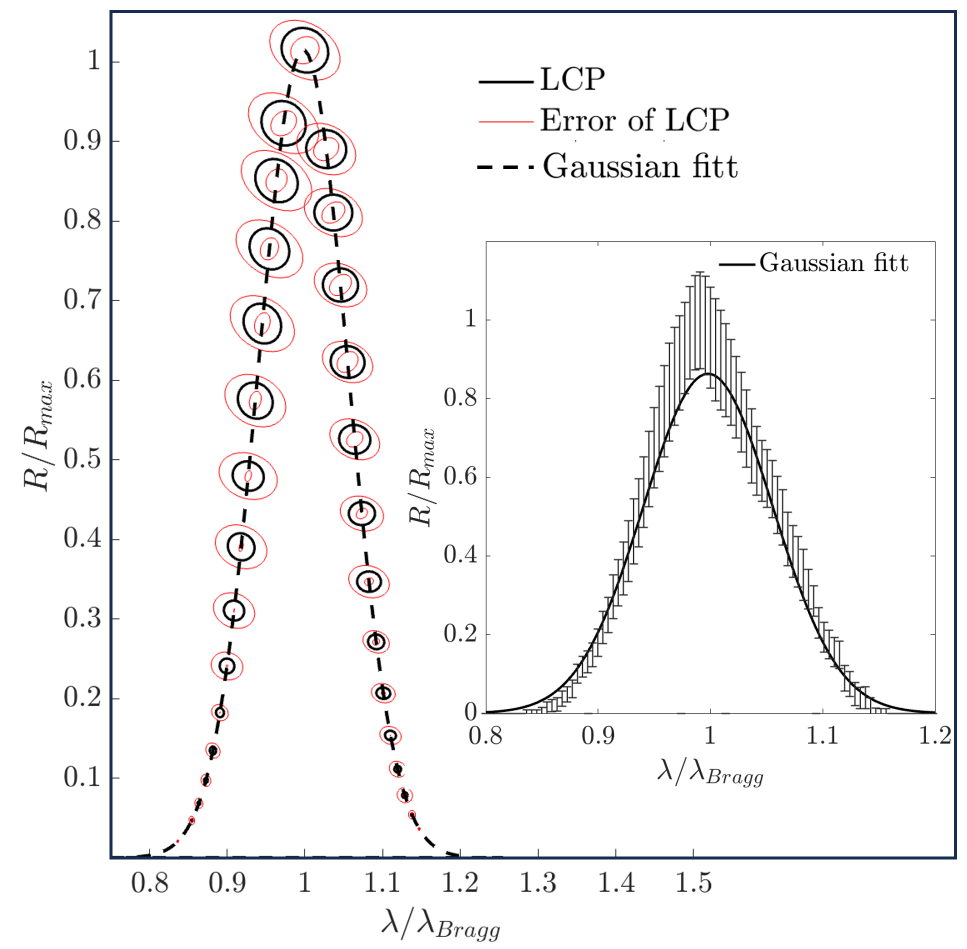}
\caption{Analysis of normalized reflectance spectra. The spectra from ten different regions across three beetle specimens are fitted to a Gaussian (an example of a Gaussian fit is shown in the inset) to determine the center wavelength, width of the spectrum and peak reflectance. The Gaussian fits are normalized to maximum reflectance and plotted as a function of scaled wavelength $\lambda/\lambda_\text{Bragg}$. These reduced and scaled spectra follow a universal curve emphasizing the self-similarity of the chiral Bragg reflectance in \textit{Protaetia speciosa jousselini}. The symbols in the main figure illustrate the state of polarization as a function of scaled wavelength.}\label{fig7}
\end{figure}

Figure~\ref{fig7} shows a more detailed analysis of the self-similarity of the chiral Bragg reflectance in \textit{Protetia speciosa jousselini}. The data from three different specimens across ten different locations is reduced by a Gaussian fit to the spectra. These reduced, Gaussian spectra, are scaled by normalizing the reflectance to the maximum reflectance (vertical axis) and dividing the wavelength by the central wavelength $\lambda_\text{Bragg}$. The self-similarity of the structure is evident from the universal curve in Fig.~\ref{fig7}. 

The chiral response of \textit{Protetia speciosa jousselini} follows from the measured elliptical state of polarization and is indicated by the ellipses in Fig.~\ref{fig7}. The elliptical polarization is consistent with a bulk left-circular polarized reflector and a Fresnel reflection from the surface.  As can be seen in the figure the state of polarization can be described as a single function of the scaled wavelength $\lambda/\lambda_\text{Bragg}$. 

For unpolarized light propagating along the axis of a continuous chiral structure, only circularly polarized light that matches the chirality is reflected. In this case, reflection of left-handed circularly polarized light $(R_\text{LC})$ occurs at the central wavelength, and the intensity of this reflection diminishes exponentially as a function of thickness  $L$~\cite{Takahashi}:

\begin{align}
R_\text{LC} &\propto 1-\exp\left(-\frac{L}{\ell_\text{B}}\right)
\label{eq4}
\end{align}
where $\ell_\text{B}$ is the characteristic Bragg length, which is expressed in terms of the photonic strength $\psi_C$ and pitch $\Lambda$ as

\begin{align}
\ell_\text{B} &= \frac{2\Lambda}{\pi \psi_C}
\label{eq5}
\end{align}

\begin{figure}[h]
\centering
\includegraphics[width=0.7\columnwidth]{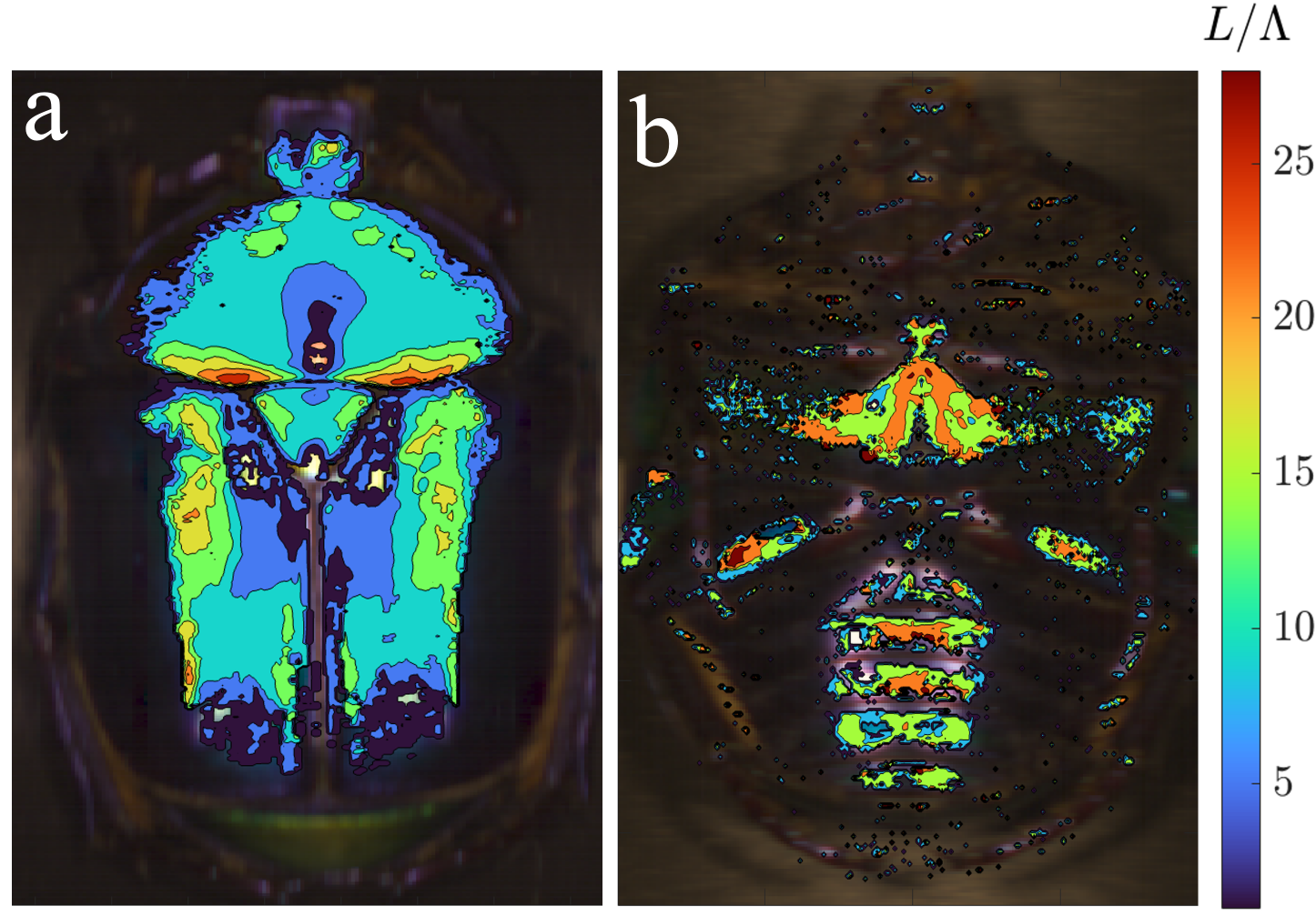}
\caption{Variation in the thickness of the chiral Bragg structure expressed for the top (a) and bottom (b) of the beetle abdomen as the number of periods $L/\Lambda$.}\label{fig8}
\end{figure}

We assume that the structures have a constant epicuticle refractive index constrast and no absorption. With these assumptions the variation in reflectance from different parts of the beetle's exocuticle stems from differences in the length of the chiral structure, denoted by $ L $, in terms of $\Lambda$: 

\begin{align}
\frac{L}{\Lambda} &= -\frac{2}{\pi \psi}\ln\left({\frac{1-R_\text{max}}{1-R_\text{min}}}\right)
\label{eq6}
\end{align}

where $R$ refers to the reflectance and $R_\text{max}$ refers to the maximum left-handed circularly polarized reflectance. The minimum reflectance $R_\text{min}$ should be thought of as a correction due to a combination of imperfections in the optical setup and background reflection, e.g. due to an imperfect anti-reflection coating. Equation~\ref{eq6} thus expresses the number of layers in the Bragg stack obtained from the non-unity reflectance. The resulting number of layers in the beetle’s exocuticle as illustrated in Fig.~\ref{fig8}.

From our measurements we determine the thickness of the Bragg layers in the exocuticle which may translate to thickness variations in the endocuticle that is important for mechanical strength of the beetle. It should be noted that the beetle \textit{Protaetia speciosa joussilini} is, despite it's colorfull appearance, a relatively simple chiral photonic structure with a single chiral Bragg reflector. The photonic strength can be used as a measure of the birefringence of the consituent chitin fibers.  Compared to other scarab beetles the relative width of the reflectance, i.e. the photonic strength is small compared to species that appear gold (\textit{Chrysina aurigans}) or silver (\textit{Chrysina limbata})~\cite{Campos-Fernandez2011}. These species get their structural color from multiple stacked chiral Bragg layers. Analysing these more complex structures in terms of photonic strength and scaling is non-trivial.

\section{Conclusion}\label{sec13}
The study of chiral structures in nature, in particular the beetles \textit{Protaetia speciosa jousselini}, offers  exciting opportunities for advancing our understanding of photonic materials. The combination of hyperspectral imaging and polarization-resolved measurements provides a powerful framework for exploring the optical properties of chiral nanostructures. By using hyperspectral polarization-resolved imaging, we can obtain comprehensive data on the photonic structure. 

These data provide insight in the origin of the homochiral optical response in nature in general and beetles in particular. A detailed spectrally and polarization resolved imaging technique is required to explore the classification of beetles across the phylogenetic tree: a classification should include narrow band spectral features at specific locations as well as wavelengths outside the visible spectrum. Such classifications can help to confirm that a macroscopic chiral response originates from homochirality on a molecular level, i.e. of chitin molecules.

The left-circularly polarized reflectance of \textit{Protaetia speciosa jousselini} is self-similar, which implies that both the reflectance spectrum and the state-of-polarization can be plotted as a universal curve as a function of the wavelength normalized to the central wavelength. At the central wavelength a near 100\% circular polarization response is found. The degree of circular polarization decreases away from the central wavelength and the fraction of circular polarization changes sign at the edges of the gap. We interpret this sign change as an effect caused by Fresnel reflection of a non-prefect anti-reflection coating of the chiral structure at that wavelength. In addition, effects as a function of angle may cause a change in sign of the circular polarization. We stress that neither of these effects contradict the relation between the macroscopic  homochiral response and the homochirality of chitin.  

We find a relative chiral stopgap width, i.e. a photonic stength paramater $\psi_C \approx 0.14$. Within this framework the thickness $L/\Lambda$ of the Bragg stack can be deduced from the measured reflectance and we find variations in this thickness across the beetle. These insights can inform the design of advanced photonic devices, bridging the gap between natural and synthetic systems. 

\begin{backmatter}
\bmsection{Funding}
This research is part of the Dutch National Science Agenda under project NWA.1306.22.017 and is made possible by financial support of the Dutch Research Council (NWO). AvdM is funded by the FCT – Fundação para a Ciência e a Tecnologia under contract number DL57/2016/CP1440/CT0009.

\bmsection{Acknowledgements}

We acknowledge Daan Lytens, Ayla Nieuwenhuijs, Time Reisinger and Naor Scheinowitz for fruitful discussions. This research is part of the Dutch National Science Agenda under project NWA.1306.22.017 and is made possible by financial support of the Dutch Research Council (NWO). AvdM is funded by the FCT – Fundação para a Ciência e a Tecnologia under contract number DL57/2016/CP1440/CT0009.

\bmsection{Disclosures} The authors declare no conflicts of interest.

\bmsection{Data availability} Data underlying the results presented in this paper are not publicly available at this time but may be obtained from the authors upon reasonable request.

\end{backmatter}
\bibliography{hyperspectral}






\end{document}